\documentstyle[twocolumn,prl,aps,epsfig]{revtex}

\begin{document}
\renewcommand{\thefootnote}{\fnsymbol{footnote}}
\sloppy

\newcommand \be{\begin{equation}}
\newcommand \bea{\begin{eqnarray} \nonumber }
\newcommand \ee{\end{equation}}
\newcommand \eea{\end{eqnarray}}
\newcommand{\rar}{\rightarrow}
\newcommand{\eq}{equation}
\newcommand{\eqs}{earthquakes}
\newcommand{\rp}{\right)}
\newcommand{\lp}{\left(}

\twocolumn[\hsize\textwidth\columnwidth\hsize\csname
@twocolumnfalse\endcsname

\title{Acoustic radiation controls dynamic friction: \\
Evidence from a spring-block experiment}

\author{Anders Johansen$^1$ and Didier Sornette$^{1,2,3}$}
\address{$^1$ Institute of Geophysics and Planetary Physics\\
University of California, Los Angeles, California 90095}
\address{$^2$ Department of Earth and Space Science\\
University of California, Los Angeles, California 90095}
\address{$^3$ Laboratoire de
Physique de la Mati\`ere Condens\'ee\\
CNRS UMR6622 and Universit\'e des
Sciences, B.P. 70, Parc Valrose\\ 06108 Nice Cedex 2, France}
\date{\today}
\maketitle

\begin{abstract}
Brittle failures of materials and earthquakes generate acoustic/seismic
waves which
lead to radiation damping feedbacks that should be introduced in
the dynamical equations of crack motion. We present direct
experimental evidence of the importance of this feedback on
the acoustic noise spectrum of well-controlled spring-block sliding experiments
performed on a variety of smooth surfaces. The full noise spectrum
is quantitatively explained by a simple noisy harmonic
oscillator equation with a radiation
damping force proportional to the derivative of the acceleration,
added to a standard viscous term.
\end{abstract}
\vspace{5mm}
]

\narrowtext

The science of solid friction has a long history,
dating back in the western world
to the geometrical work of Leonardo di Vinci, continuing with
the empirical Amontons' law two centuries later and Coulomb's investigations
of the influence of sliding velocity on friction in the 18'th century. Some
40 years ago, R. Feynmann in his famous lectures stated that ``Friction is a
very complicated matter... and in view of all the work that has been done it
is surprising that more understanding of this phenomenon has not come about.''

Only three decades ago was it recognised that friction plays a role in the
mechanics of earthquakes \cite{BB66} and it was proposed that stick-slip
as observed in friction \cite{leben} was relevant to earthquake dynamics.
Numerous laboratory experiments have been carried out to identify the relevant
parameters controlling the solid friction and hence the stick-slip behaviour
\cite{PT96}. Low velocity (below $\approx 1$ cm/s) experiments have
established that solid friction is a function
of both the velocity of sliding and of one or several state parameters,
characterising the true surface of contact \cite{statevar}. These so-called
Ruina-Dieterich laws now constitute the basic ingredients in most models
and numerical elasto-dynamic calculations directed at understanding
the process of rupture and earthquake nucleation.

A well-known and serious limitation of these calculations based on laboratory
friction experiments is that the friction laws used have been determined under
steady-state sliding conditions using velocities of no more than
$\approx 1$~cm/s (and often much less), {\it i.e.}, considerably below the
sliding velocity of m/s  occuring during an earthquake.
Thus, one may ask whether it is correct to extrapolate these laws and their
velocity weakening dependence to higher velocities relevant to \eqs ? Such
considerations become all the more relevant when one examines the underlying
physical mechanisms of the friction laws. At low
velocity, effects such as hysteretic elastic and plastic deformations at
the scale of roughness asperities seem to play a dominant role
\cite{lowvel}. At larger velocities, new mechanisms come into play.
Collisions between asperities and transfer of momentum between the
directions parallel and perpendicular to the motion are potentially
important mechanisms \cite{collision}. Recently, Tsutsumi and Shimamoto
\cite{TS9697} have performed friction measurements
on rotating cylindrical samples at velocities up to $1.8$~m/s and for slips
of several tens of meters. Their results indicate
a change of regime from velocity weakening to velocity
strengthening at large velocities. This is confirmed by
3-d numerical simulations performed in a regime of velocities
of meters to tens of meters per
second \cite{Maveyraud}, reflecting
the increasing strength of vibrational damping.

Radiation damping is well-documented in electromagnetism
\cite{Jackson,Electro} and nuclear physics. Surprisingly, the
corresponding mechanism of dynamic friction due to
radiation of phonons or seismic waves has received little
attention (see however \cite{sokoloff}), notwithstanding its large 
potential impact on the dynamics. For
instance, it is well-known that Burridge-Knopoff spring-block models
does not recover the correct elasto-dynamic
continuous limit but rather lead to a Klein-Gordon equation with a mass term
implying finite range interactions \cite{Carlson}. This problem can be
addressed by adding a viscous damping accounting for radiation losses
with an amplitude finely tuned to the critical damping value
\cite{Knopoffdamping}.
The influence of ultrasounds on crack dynamics in brittle
materials has been demonstrated
by using both the natural sound emitted by the propagating crack
and an artificially generated ultrasound burst \cite{Boudet}.
Although the acoustic energy is only $5\%$ of the energy needed to
propagate the crack, the presence of sound waves in the specimen strongly
modifies the fracture dynamics because the sound interacts with the crack tip.

Here, we re-analyse the high-frequency part of the power spectrum from 
an experimental investigation of stick-slip in dry metallic friction 
\cite{speciale}. We focus on the dynamic friction in the high velocity regime,
{\it i.e.}, slip  velocities ranging up to $v_{max} \approx 0.35$ m/s,
which provides a direct demonstration of the role and nature of radiation 
damping feedback on the dynamics. Other aspects of these experiments have 
been reported elsewhere \cite{speciale,prewear}, but a satisfying explanation 
for the high velocity behaviour could not be proposed at that time.

\begin{figure}
\epsfig{file=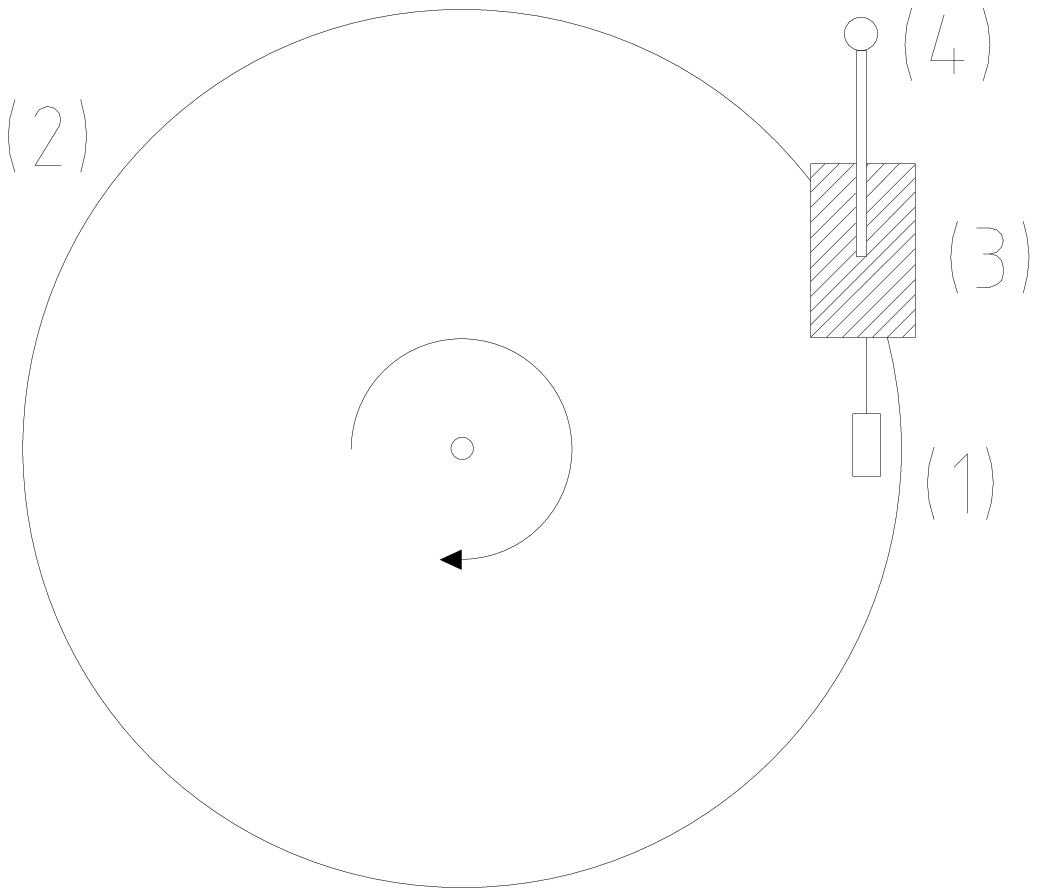, width=8.6cm,height=5cm,bbllx=-70,bblly=0,bburx=410,bbury=290}
\caption{\protect\label{setup1} Top view of the experimental setup. A 
small steel block (1) is placed on a rotating steel table (2) and via a thin 
plastic rod attached to a steel spring. The steel spring is placed inside a 
metal box (3) kept fixed in the laboratory frame (4). The metal box (3) forms 
a common shield with the electronic measuring 
devices and the battery (not shown) powering the Wheatstone bridge containing 
the strain-gauges thus reducing electro-magnetic noise. Details of the setup 
contained in the metal box (3) is shown in figure \protect\ref{setup2}.}
\epsfig{file=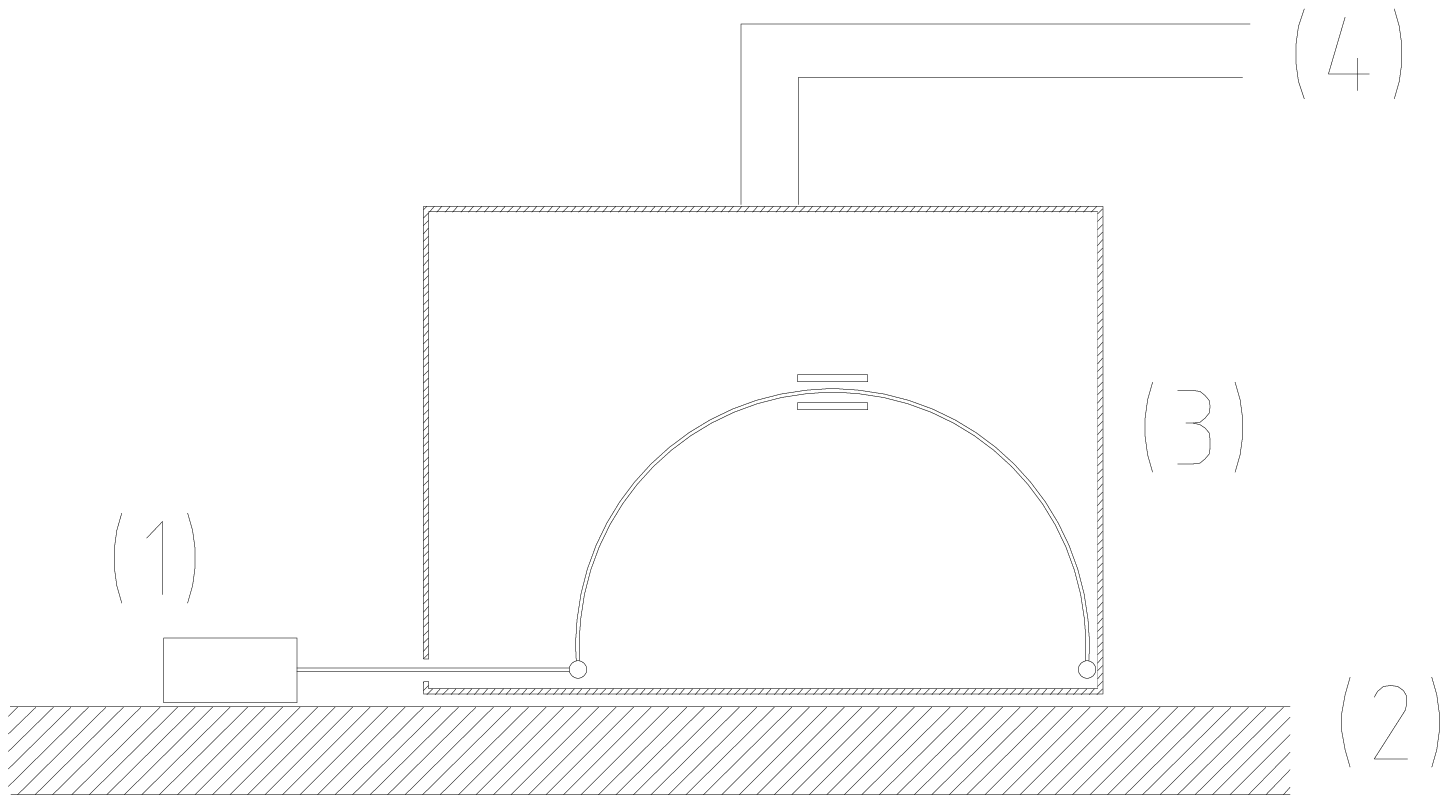, width=8.6cm,height=6cm,bbllx=0,bblly=0,bburx=480,bbury=300}
\caption{\protect\label{setup2}Side view of the metal box containing the 
spring and the strain gauges glued on each side of the spring.} 
\end{figure}

Details of the experimental setup can be found in \cite{speciale,prewear} 
and here we will just repeat the essential features, see figures \ref{setup1}
and \ref{setup2}. The experiment consisted of a small steel block (1) placed 
on a rotating steel table (2) and via a thin plastic rod attached to a steel 
spring inside a metal box (3), see figure \ref{setup1} and caption. Other 
metal 
surfaces were used and gave similar results. The other end of the steel spring 
was kept fixed in the laboratory frame of reference (4). As the surface of 
the table moved, the spring became elongated until the spring force exceeded 
the friction force on the block and the spring contracted until the friction 
force again exceeded the spring force and so forth. The mass of the block was 
$28.5$ g and the spring constant $22.4$ N/m giving a natural frequency of the
spring and block $\omega_0 = \sqrt{k/m} \approx 28 ~ \mbox{s}^{-1}$. The 
elongation of the spring 
was measured using strain gauges mounted on each side of the spring, see figure
\ref{setup2}, and placed in a Wheatstone bridge. The use of batteries both for 
the bridge and the pre-amplifier as well a common shield reduced 
electromagnetic noise significantly and the over-all signal-to-noise ration 
was better than 1000:1, corresponding to an elongation of the spring of 
$~5\mu$m to $~5$ mm or more.

When the block moves relatively to the surface of the table, there
are many collisions between the asperities of the two surfaces. Now, 
the asperities typically have a spacing of $10-100$ microns, which means
that the block (on average) will experience thousands of collisions between
asperities per second. The motion of the block can thus be modeled as a 
``noisy'' damped harmonic oscillator with the following equation of motion 
for the position $x(t)$ of the block
\be\label{blockeq}
\ddot{x} + c_{rad}\stackrel{{\mbox{...}}}{x} + c_{vis} \dot{x} + \omega_0^2
x =\eta (t),
\ee
where $\eta (t)$ is a ``white'' noise term, {\it i.e.}, $\left< \eta (t) 
\right> =0$ and $\left<
\eta (t) \eta (t') \right> =b \delta (t-t')$ accounting for the stochastic
motion of the asperities. We introduce two damping terms. The non-standard
$c_{rad} \stackrel{{\mbox{...}}}{x}$ term is the Abraham-Lorenz expression
for the
first-approximation of the direct reaction force due to radiational damping.
A viscous friction $c_{vis} \dot{x}$ must also be present in order to produce
a friction force at constant velocities. It includes the effect of
all other friction mechanisms, including a renormalisation of the
Abraham-Lorenz term at the scale of individual asperities that undergo
acceleration/deceleration even under
constant block velocity. As we shall see below, the radiative term
dominates completely above the characteristic frequency $\omega_0$.

It is useful to recall the derivation of the radiation damping reactive force
$F_{rad}=c_{rad} \stackrel{{\mbox{...}}}{x}$. We start from
the general expression for the power $P(t)$ radiated by this
element. By Galilean invariance, $P(t)$ must be zero if the velocity
$\dot{x}$ is constant and becomes
non-zero when the acceleration $\ddot{x}$ is non-zero. Assuming
analyticity and symmetry under $\ddot{x} \to - \ddot{x}$
and performing a Taylor expansion in powers of $\ddot{x}$, we get
the leading term as
\be
P(t) = m_e \tau \ddot{x}^2~.
\label{hfjjkd}
\ee
The quadratic dependence of the radiated power by a small acceleration
element is so general that it applies to any physical problem involving an
accelerating body coupled to a wave. $m_e$ is the mass of the element and
$\tau$ is a characteristic time proportional to $R/c$ where $R$ is the
typical linear size of the element and $c$ is a wave velocity.
Expression (\ref{hfjjkd}) recovers the Larmor power
formula for
the electromagnetic radiation of an electric charge. It also describes the
acoustic radiation
from an accelerating volume element, the fluid gravity waves radiated from
an accelerating
surface distortion or even the gravitational waves from an accelerating
black hole.

In order to obtain the expression of the reaction force $F_{rad}$ due to
radiation, we follow
Jackson \cite{Jackson} and view, by the requirement of energy conservation,
 the radiated power $P(t)$ as minus the work per unit time
of $F_{rad}$\,: $\int_{t_1}^{t_2} dt ~F_{rad} ~\dot{x} = -
\int_{t_1}^{t_2} dt~P(t)$.
Integrating the r.h.s. by part and neglecting the boundary term (which are
zero for
periodic motion), we get the Abraham-Lorenz expression for the radiative force
\be
F_{rad} = m_e \tau\stackrel{{\mbox{...}}}{x}~.
\label{hfjkfkx}
\ee
As a first-order approximation, it must be replaced by an integro-differential
equation when radiation damping becomes the dominant term in the dynamics.

The consequence of this result is dramatic. For a given oscillatory
amplitude, the radiative damping is proportional to $\omega^3$ compared to
the usual $\omega$ for viscous damping. This corresponds to a weaker damping
at low frequency and a more efficient effect at large frequencies (but still
sufficiently small so that the wavelength remains larger than the source size).

In (\ref{blockeq}), a renormalised coefficient $c_{rad}$ is used instead of
$m_e\tau/m$ to account for the collective behaviour summed over all asperities.
We can estimate it theoretically by transforming expression (\ref{hfjjkd})
in the Fourier domain as $P_{\omega} = c_{rad} m \omega^2 V^2$, where $m$ is
the block mass and $V$ the particle velocity of the sound wave
generated by the moving mass. We then equate this expression to the
radiated power of $N$ coherent asperities of radius $R$ and contact
pressure $p$
given by $1.2 N^2 \omega^2 (\pi R^2 p)^2 / \rho c^2$ \cite{Miller}, where
$\rho \approx 7.8$g/cm$^3$, $c \approx 5800$m/s for steel and the coefficient
$1.2$ is for Poisson ratio of $1/4$ and varies slowly. Notice that
$N \pi R^2 p \approx mg$, assuming that the contacts are represented by discs
of radius $R$, where $g=9.8~$m/s$^2$ is the earth acceleration. This
approximation
assumes that there is no significant additional inertial pressure due to
vertical motion of the block other than
its static weight on the contact asperities. This leads
to the following simple expression\,:
\be
c_{rad} \approx {1.2 \over \rho m} \biggl( {mg \over Vc}\biggl)^2~.
\label{dqmqmmq}
\ee
All parameters in (\ref{dqmqmmq}) are known except the particle wave velocity
$V$, which can be determined from our fit to the spectrum.

\begin{figure}
\begin{center}
\epsfig{file=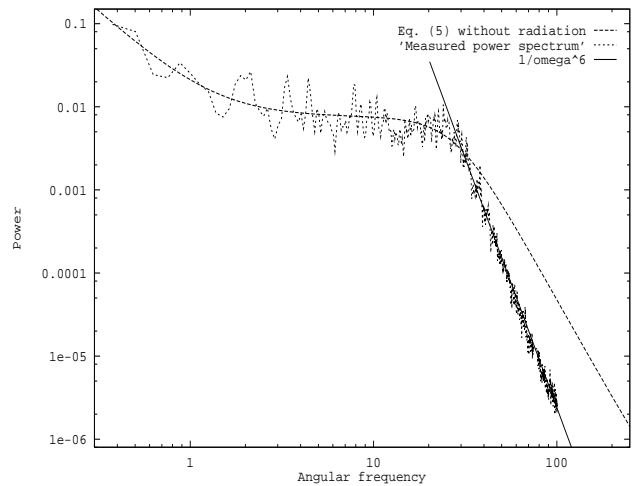, width=8.6cm,height=7cm}
\caption{\protect\label{powfig} The power spectrum of the motion of the
block in the experimental setup shown in figures \protect\ref{setup1}
and \protect\ref{setup2}. The fit is \eq \ (\protect\ref{poweq}) with a
$a/ \omega^2$-term added to account for the apparent drift and with 
$c_{rad} = 0$ and $\omega_0 = 28 \mbox{s}^{-1}$. The high frequency tail of 
the spectrum has been fitted with a pure $\omega^{-6}$-decay.} 
\end{center}
\end{figure}

The power spectrum corresponding to (\ref{blockeq}) is
\be\label{poweq}
S ( \omega ) = \frac{2b}{\lp \omega^2 -  \omega_0^2 \rp^2 +
\omega^2\lp c_{rad}  \omega^2 - c_{vis} \rp^2}~,
\ee
leading to an $\sim 1/ \omega^6$-decay of the power spectrum for 
frequencies larger than the natural frequency $ \omega_0 \approx 28 ~
\mbox{s}^{-1}$. Figure \ref{powfig} shows the power spectrum measured 
experimentally in the regime where the block 
was constantly sliding. Note the corner-frequency (or shoulder) at 
$\omega_0 \approx 28~\mbox{s}^{-1}$. There appears to be a $1/\omega^2$ 
background 
below this corner frequency. This is most likely an artifact of making a 
finite time measurement thus creating an illusion of a slow constant 
drift in the signal. Indeed, if $x(t) \rar x(t) + a t$ then
$\tilde{x}(\omega ) \rar \tilde{x}( \omega ) + a/i\omega$ (for periodic
boundary conditions) resulting in the addition of a $a^2/ \omega^2$-term
to \eq \ (\ref{poweq}). 

In figure \ref{powfig} is also shown a fit to the data with \eq \ (\ref{poweq})
with this correction term added and with only viscous damping, {\it i.e.},
($c_{rad}=0$). It emphasises the importance of the radiation term to account 
correctly for high frequency tail of the spectrum. In figure \ref{powfig} we
also show a fit of the high frequency tail of the spectrum with the predicted
$1/\omega^6$-decay from the radiation term. The agreement between the data and
the prediction is excellent. The fit of the entire spectrum with both damping 
terms non-zero in \eq \ (\ref{poweq}) is found to be highly unstable due to 
the complete dominance of the radiative damping compared to the viscous 
damping indicating that most of the useful high-frequency spectrum
shown in figure \ref{powfig} is controlled by the novel radiation damping term.

From the fit of the high frequency tail of the spectrum we get $c_{rad} 
\approx 0.02~$s. Using (\ref{dqmqmmq}) we obtain an estimation
$V \approx 25 ~\mu$m/s for the wave particle velocity giving an acoustic 
pressure $\rho c V \approx 1100~$Pa corresponding to very energetic local 
sources.

This finding may illuminate the so-called ``cut-off'' frequency problem
\cite{Aki}, on the source acceleration spectra of moderate to strong 
Californian earthquakes. The unexpected result is that the seismic spectrum 
falls off very fast beyond a frequency attributed to a characteristic size 
of cohesive fault size or to scale-length of heterogeneities of the fault 
plane. Our results may quantify this phenomenon by confirming the role of 
the radiation by accelerating asperities and specifies the quantitative 
shape of the spectral fall-off. However, earthquakes faults have a thick
layer of fault gouge between the sliding rock blocks which might affect
the magnitude of the radiation term. In order to clarify this issue, further 
experimental work with such an intermediate layer should be performed.

In conclusion, we have presented a new analysis of high sensitivity 
measurements of noise spectra in spring-block experiments \cite{speciale}
with sliding velocities in the range $\approx 0.35$ m/s. We have shown 
that the spectrum exhibit an approximate $1/\omega^6$-decay
for frequencies larger than the natural corner value
$\omega_0 = \sqrt{k/m}$. This $\omega$-dependence can be rationalised
simply by the generic Abraham-Lorenz radiation damping law, with a
reactive force proportional to $\stackrel{\mbox{...}}{x}$, generated by
the radiation of sound waves due to collisions of asperities.
We expect that this finding extends to the slip-stick regime for which
measurements of the two-point correlations of successive slip
characteristics, slip distance and time of slip were found to be very weak
\cite{prewear} indicating highly nonlinear dynamics.
The implication of this finding for rupture and
earthquake modelling is of great potential impact. Our results suggest that the
Ruina-Dieterich
friction laws cannot be extended to the high velocity regime relevant for
earthquakes and many cases of rupture. Future works include the generalisation
of the acoustic
Abraham-Lorenz radiation damping beyond the first-order approximation and the
derivation of a generalised
Fluctuation-Dissipation theorem relating the noise amplitude $b$
to the damping coefficients $c_{rad}$ and $c_{vis}$.

\acknowledgements

We acknowledge useful discussions with K. Aki and the help of K. Grum-Schwensen
with the figures.


\begin{thebibliography}{}


\bibitem{BB66} W. R. Brace and J.D. Byerlee, Science 153, N3739, 990 (1966).

\bibitem{leben}  F.P. Bowden and L. Leben, Proc. Roy. Soc. (London) A
169, 371 (1939).

\bibitem{PT96} B.N.J. Persson and E. Tosatti, eds., {\em Physics of sliding
friction}, NATO ASI Series, Kluwer Academic Publishers, Dordrecht (1996).

\bibitem{statevar} See for instance:
W.F. Brace, Tectonophysics 14, 189-200, (1972).
J.H. Dieterich, J. Geophys. Res. 77, 3690-3697 (1972);
Pure and Applied Geophysics 116, 790-806 (1978);
J. Geophys. Res. 84, 2161-2168, (1979);
Tectonophysics 211, 115-134 (1992).
A. Ruina, J. Geophys. Res. 88, 10359-10370 (1983).
S.J.D Cox in  {\it Deformation Mechanisms, Rheology and Tectonics},
Knipe et Rutter,
eds., vol. 54, 63-70, Geological Society Special Publication (1990).
N.M. Beeler {\it al.}, J. Geophys. Res. 101, 8697-8715 (1996).
N.M. Beeler {\it al.}, Geophys. Res. Lett. 21, 1987-1990 (1994).
C.H. Scholz, Nature 391, N6662, 37-42 (1998). G. Zheng and J. R. Rice,
Bull. Seism. Society Am., Vol 88, No. 6 1466-1483 (1998).

\bibitem{lowvel} See for instance:
F.P. Bowden and D. Tabor, {\em The Friction and Lubrication of Solids},
Oxford University Press (1954)
H.J Jensen {\it al.}, J. Phys. I France 3, 611-623 (1993).
J. Dieterich and B.D. Kilgore, Pure and Applied Geophysics 143, 283-302 (1994).
C. Caroli and P. Nozi\`eres in {\it Physics of Sliding Friction}, Persson
et Tosatti,
eds., Kluwer Academic Publishers (1996).
A. Tanguy, and P. Nozi\`eres, J. Phys. I France 6, 1251-1270 (1996).
A. Tanguy and S. Roux, Phys. Rev. E 55, 2166-2173 (1997).
C. Caroli and B. Velicky, J. Phys. I France  7, 1391-1416 (1997).
L. Bocquet and H.J. Jensen, J. Phys.  I France 7, 1603-1625 (1997).

\bibitem{collision} See for instance:
J.A.C. Martin and J.T. Oden, Comp. Meth. Appl. Mech. Eng. 52, 527 (1985);
J.A.C. Martin, J.T. Oden and F.M.F Sim\~{o}es,
Int. J. Eng. Sci. 28, 29 (1990).
J. Lomnitz-Adler, J. Geophys. Res. 96, 6121 (1991).
D. Pisarenko and P. Mora, Pure and Applied Geophysics 142, 447 (1994).

\bibitem{TS9697}
A. Tsutsumi and T. Shimamoto, J. Geol. Soc. Japan 102, 240-248 (1996).
Geophys. Res. Lett. 24, 699-702 (1997).

\bibitem{Maveyraud} C. Maveyraud {\it al.}, J. Geophys. Res. (cond-mat/9809213)

\bibitem{Jackson} J.D. Jackson, Classical Electrodynamics, 2nd ed. (John
Wiley and Sons, New York, 1975), p. 783.

\bibitem{Electro} F.V. Hartemann and N.C. Luhmann, Phys. Rev. Lett. 74,
1107 (1995).

\bibitem{sokoloff} J.B. Sokoloff, Phys. Rev. Lett. 71, 3450 (1993).
M.S. Tomassone {\it et al.}, Phys. Rev. Lett. 79, 4798 (1997).


\bibitem{Carlson} J.M. Carlson {\it et al.}, Rev. Mod. Phys.
66, 657-670 (1994).

\bibitem{Knopoffdamping} H.J. Xu and L. Knopoff, Phys. Rev. E 50, 3577
(1994).

\bibitem{Boudet} J.F. Boudet and S. Ciliberto, Phys. Rev. Lett. 80, 341
(1998).

\bibitem{speciale} A. Johansen,
\newblock Dynamics and Statistics of a Stick-Slip
Experiment. Master Thesis, Niels Bohr Inst. (Mar. 1993).

\bibitem{prewear} A. Johansen, P. Dimon, C. Ellegaard, J.S. Larsen and H.H. 
Rugh, Phys. Rev. E 48, 4779 (1993). 
A. Johansen, P. Dimon and C. Ellegaard , Wear 172, 93 (1994).

\bibitem{Miller} G.F. Miller and H. Pursey, Proc. R. Soc. London, Ser. A 233,
55 (1955); A.M. Stoneham and A.H. Harker, Wear 80, 377 (1982).

\bibitem{Aki} A. Papageorgiou and K. Aki, Bull. Seism. Soc. Am. 73, 693 (1983);
ibid 73, 953 (1983); PAGEOPH 123, 353 (1985);
A. Papageorgiou, Bull. Seism. Soc. Am. 78, 509 (1988).


\end{thebibliography}
\end{document}